# Enhancing Business Process Execution with a Context Engine

Christian Janiesch, Faculty of Business Management and Economics, University of Würzburg, Würzburg, Germany

Jörn Kuhlenkamp, TU Berlin, Berlin, Germany

## 1 Introduction

Business processes represent transactions internal to or between companies, which take place over a certain amount of time. Business processes do not necessary have to be supported or be executed by a business process management (BPM) system. If they do, however, they have the benefit of coordination, which greatly reduces the effort of the process owner to keep track of unclaimed tasks, sequence, logging and so forth.

At the time of execution, business processes are commonly instantiated by the BPM system with all relevant parameters to distribute tasks correctly. The Workflow Management Coalition (WfMC) (1999) defines these parameters, workflow relevant data*,* as "data that is used by a Workflow Management System to determine the state transitions of a workflow instance, for example within pre- and post-conditions, transition conditions or workflow participant assignment."

However, long running processes may require that these parameters, the process's context, are updated or extended during execution and that the flow of the process can be adapted. For example, fluctuations in exchange rates, change of weather patterns, or traffic congestions can have an impact on logistics processes and change their profitability or lead to failed instances (e.g., late deliveries). However, not all parameters may be known at initialization. While WfMC's definition does not explicitly exclude this understanding, its aim is to define data internal to the BPM system (Workflow Management Coalition, 1999).

Consider for example a logistics process of the delivery of a spare part for mining machine. Once the need for a spare part has been signaled, the machine provider may have only a certain timeframe to replace the part due to the current service level agreement (SLA). If the delivery is bound to delay beyond that SLA due to a context factor (e.g., weather: road washed away, delay at the port), the machine provider may elect to use a different means of transportation to stay within the timeframe allotted by the SLA. This requires the adaptation of running processes based on context data.

Having recognized this issue, in recent years the concept of context-aware information systems has gained more traction (Wieland *et al.*, 2007). The aim is to provide some sort of context to the execution of applications. However, in the area of BPM most research has touched only aspects of context-awareness, such as process adaptation, process modeling, or managerial aspects of processes (Hallerbach *et al.*, 2008, Rosemann *et al.*, 2006). The description or implementation of an architecture to capture and process context outside the process's control flow has been absent from the discussion or has only been focused on a mostly static part of model configuration (Hallerbach *et al.*, 2008) or a closed environment scenario, for example of a smart factory (Wieland *et al.*, 2011).

This poses the following research question:

*"How can we model the context of a business process, and how can we realize a system architecture for context management and dynamic process adaption in support of these models?"*

In the following, we propose to use a context engine to add the missing flexibility to BPM using business rules and complex event processing (CEP) technology. While rules engines are typically tightly integrated with BPM systems, a CEP engine is considered a separate system. CEP provides technology to process a large amount of events (Luckham, 2002), which can signify context changes in real-time.

We describe an architecture to manage context and the context-aware execution of processes, which involves a process execution environment, a rules engine to take or suggest decisions, and a context engine to maintain and provide context for the former two systems. In this paper, we explicitly focus on an architectural view of the application and do not go into details on possible methods to mine and analyze context change.

The paper is structured as follows: In Section 2, we elaborate on related work to context and context modeling and provide the baseline for a context-aware architecture for BPM. Section 3 comprises the formal scope of our context model. In Section 4, we introduce the architecture and capabilities of the proposed system. In Section 5, we describe the general process of context-driven process adaptation and decision support and give an example. The paper closes with a discussion, summary and outlook.

# 2 Related Work

### 2.1 Business Process Management Systems, Rules Engines, and Complex Event Processing

Processes are commonly seen as activities or tasks executed within or across enterprises or organizations (Object Management Group Inc., 2013). Thereby, activities may be complex tasks or atomic tasks. Complex tasks do not need to be analyzed further. Atomic tasks cannot be divided any further. Each activity is either of automated or manual nature.

BPM is regarded as a collection of methods and tools for creating a common understanding of a company's process portfolio and then for managing and improving this portfolio (zur Muehlen and Indulska, 2010). The aim is to plan, to control, and to monitor intra- and inter-organizational processes with regards to existing operational sequences and structures in a consistent, continuous, and iterative way of process improvement (Becker *et al.*, 2011). A BPM system or engine allows for the definition, execution, and logging of business processes. The modeling of business processes aims at condensing complex process descriptions into an accessible graphical representation, which can be used to simplify and discuss as well as to document and specify for implementation.

A (business) rules engine executes a set of rules to find a solution for a business problem based on some input variables. "Rules" can include rule statements, facts, priorities, mutual exclusion, preconditions or complex decision processes. While rules engines can stand alone, they often complement a BPM system. Here, the main aim is to separate business logic from the application or process model (von Halle, 2001).

The central concept in CEP is the event. In everyday use, an event is something that has occurred. In CEP, an event signifies an activity, which has happened, by recording this information in a data object. The event signifies the activity, for example a context change. Events comprise of several attributes (Luckham, 2002). CEP encompasses techniques for collecting and analyzing events in any system by gathering lower-level system events and deducing higher-level knowledge from those in real-time (Etzion and Niblett, 2011, Luckham, 2002). CEP engines execute an event processing network in which individual agents filter, match, and derive events. In this way, they can reduce the amount of events to the desired selection, correlate them with other events to find causalities or derive new complex events. In terms of derivation, CEP can translate, aggregate, split, and compose events. By translating, it can reduce the information contained by projecting it or it can increase the information by enriching the individual event. Aggregations are used to define complex events based on a set of low-level events using defined abstraction relationships. Composition merges two events stream while the split operation breaks up one event into multiple events (Etzion and Niblett, 2011).

The topic of event management for BPM has received some attention in recent years. However, publications focus either on general considerations on an overall architecture (Janiesch *et al.*, 2011, von Ammon *et al.*, 2009), the modeling of process events (Kunz *et al.*, 2010) or focusses only parts of the overall concept (i.e. local workflows Wieland *et al.*, 2011). Janiesch et al (2012) provide an example of an event-driven process analysis architecture, which goes beyond process monitoring. For an overview and challenges of event-driven BPM cf. Krumeich et al (2014).

Furthermore, Adams et al. (2010, 2007), Mundbrod et al. (2015), and Nunes et al. (2016) provide situational implementations of context-aware BPM. However, neither of them consider a context engine as a separate entity or established technology such as rules engines or CEP engines for implementation. All approaches provide a good starting point for flexible process adaptation also at run-time. Yet, their results are confined to their proprietary implementations.

### 2.2 The Notion of Context and Context-aware Information Systems

The notion of context is widely used in different research areas. Examples can be found in the areas of formal logic (Arló-Costa, 1999), knowledge representation and reasoning (Brézillon *et al.*, 1998, Sowa, 1999), computational and sociological linguistics (Clark and Carlson, 1981, Halliday, 1978), cognitive psychology (Kokinov, 1999), and BPM (Hallerbach *et al.*, 2008).

A generic definition of context reads as "the circumstances relevant to an event or fact" (Crozier, 2006). Dey and Abowd (2000) understand context as "any information that can be used to characterize the situation of an entity. An entity is a person, place, or object that is considered relevant to the interaction between a user and an application, including the user and application themselves." In the context of BPM, Rosemann et al. (2006) propose a working definition of context as "the relevant subset of the entire situation of a business process that requires a business process to adapt to potential changes in the context variables."

Thus, context is understood as the combination of all implicit and explicit circumstances, which might have an impact on the situation in which it is embedded. If information can be used to characterize the situation of a process, then it is context.

Chen and Kotz (2000) understand context-aware systems as computer systems, which offer information and services to users based on their context. Examples include tool support for the design of spatial databases and management of context models (Cipriani *et al.*, 2011) or Provop, an approach for configuring context-based process variants based on static context information (Hallerbach *et al.*, 2008). Herzberg et al. (2015) present an approach to use events of object state transitions to observe process progress but do not consider external context.

Other research in BPM has been conceptual work for the most part. It has only touched aspects of context-awareness, such as reference model configuration (Delfmann *et al.*, 2006, Rosemann and van der Aalst, 2007), context-aware process modeling (Rosemann *et al.*, 2006, Saidani and Nurcan, 2009), or the context-aware management of processes (Ploesser, 2013, Ploesser *et al.*, 2010). Exception handling is also related to what we understand as context-aware. Exception handling deals with the management or failure, expiry, unavailability, and violations apart other triggers. Hence, workflow exception patterns (Russell *et al.*, 2006) can be a basis for context-based adaptation.

To structure context, several models have been developed. Anastassiu et al (2016) develop the ORGANON method to derive relevant context. Yet, they do not provide a system architecture top operationalize this information. Gu et al. (2005) use an ontology-based approach with the help of OWL (Web Ontology Language) to model context. None of the above has been designed for or applied to BPM. Serral et al. (2014) have started to adapt OWL context representations to BPM. The reference frame of Rosemann et al. (2006) facilitates the identification and integration of relevant context into business process models. Yet, the authors do not define specific context categories but only rough layers. They do not specify a data structure. Michelberger et al. (2012) formalize a context model for BPM. However, it is static and does not allow for dynamic extensions or change.

Dey (2001) applies context to the domain of computing devices and applications. They argue that it is useful to categorize context for a better, systematic comprehension due to the broad diversity of context information. For this purpose, he defines four basic context categories: identity, location, status, and time. Apart from academic approaches, there are models such as the core components and the unified context methodology using the context driver principle (UN/CEFACT, 2009) and World Wide Web Consortium's (W3C) delivery context overview for device independence (World Wide Web Consortium, 2006). Stettner and Janiesch (2009) present a list of predefined context categories based on the above. Möhring et al. (2014) also provide an initial classification of context categories. Bettini et al. (2010) provide a thorough survey of context modeling and reasoning techniques. All of the above can be valuable input when formulating a concrete context model, which is not the aim of this publication.

Summarizing, many disciplines consider and have defined context as a relevant factor. In BPM, internal variables are used as local context in a mostly static way in commercial BPM system when initializing a process model. Approaches, which deal with dynamic context in a wider sense, are of conceptual nature or proprietary implementations.

Nevertheless, several context models exist and suggest how to structure context relevant to information systems in general. None of them has been applied to BPM. In addition, the link of these models to a technical BPM system architecture using established technology for sensing and actuating is missing.

# 3 Context Model

### 3.1 Overview

We evolve the abstract notion of context into a model to define concrete context data structures: a *context model*. The context model represents the context of a business process or business process instance. It may be part of a larger *context cloud* available to a context engine. A *context master model* carries the global context of the process model or the process definition. A *context instance model* is initialized for a process instance or a group of process instances. It can be based on this master context model, which then acts as a context template for the process instance. Context instance models can be nested for business or performance reasons.

Based on the definition of context, we argue that information in these models can be represented by *context values*. Each context value within a context model is associated to a single *context category* $c \subseteq C$ with $C$ representing the set of context categories such as place, role, or time. One category can relate to other categories.

We propose a context data structure in form of a *context intersection*. A context intersection is represented by a directed, acyclic graph $G = (V, E)$ with vertices $V$ and edges $E$. $V$ represents a subset of context categories $V \subseteq C$ that are included into the context intersection. It is related to the concept of a situation from CEP (Etzion and Niblett, 2011), which represents a certain set of context values. Cf. Figure for a UML meta model as an overview.

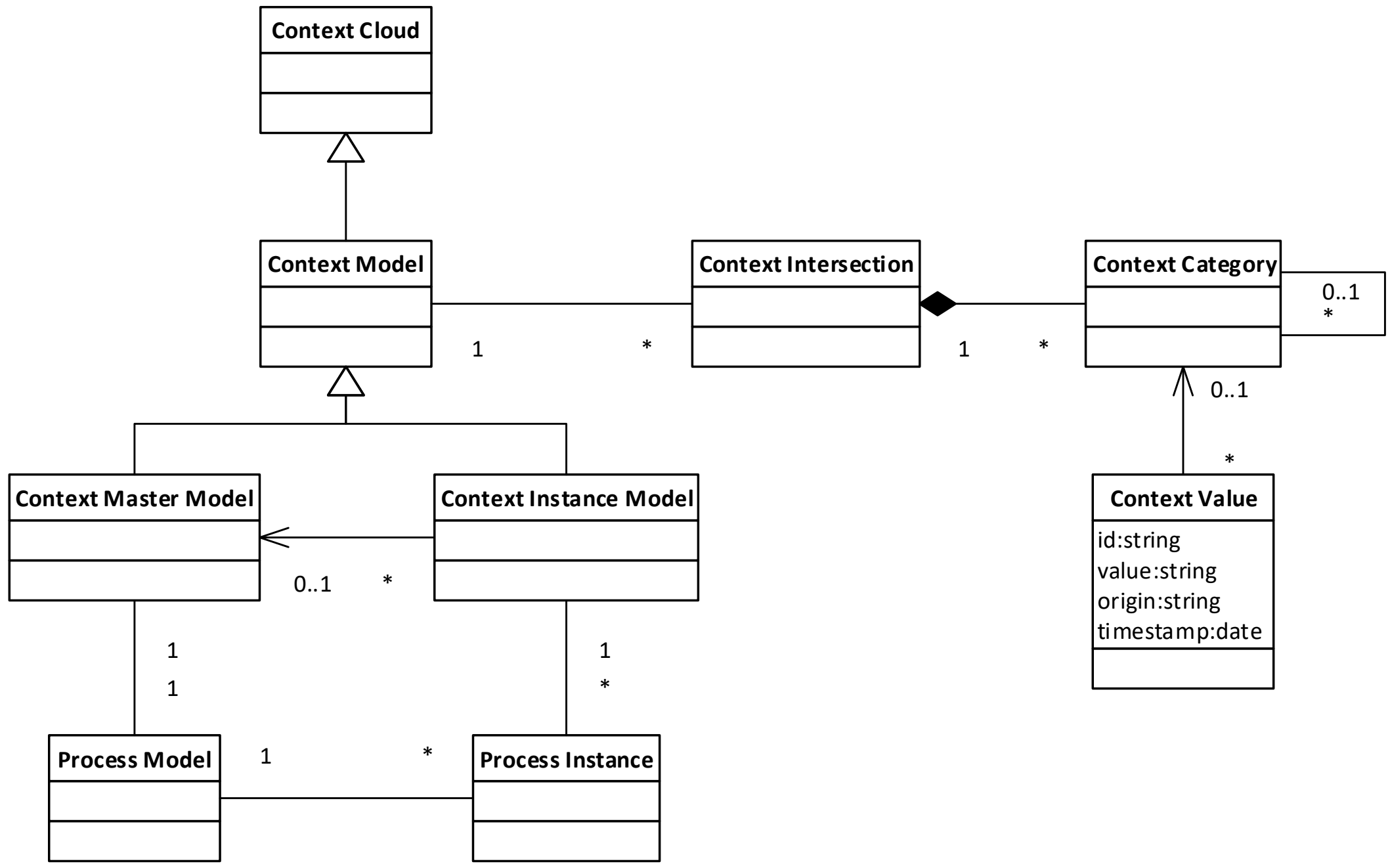

**Fig. 1.** Context Model Meta Model

Context intersections are *hierarchical*, *extensible*, and *flexible*, and context categories and their value are *time-variant* and may be in a *cause-and-effect* relationship with other context categories. In addition, there are non-functional attributes, which may be important for context processing. First, a context intersection groups multiple context categories within a hierarchy. Second, a context intersection can be extended by additional context categories before and during the execution of an instance of a business process model. Third, context intersections can differ between executions of two instances of the same business process model. Fourth, context values of one context category can be used to derive/ calculate context values of another context category. Fifth, context information is prone to decay. The older the information is, the less accurate and relevant it is towards the current situation. The five characteristics of context intersections are non-exhaustive and discussed in detail in the following. Cf. Figure 2 for an example of two hierarchical context at configuration steps $T$ and $T+1$ with the latter containing updated context values (weather) and extended context categories (packaging method, shipping method) in an example from the logistics sector.

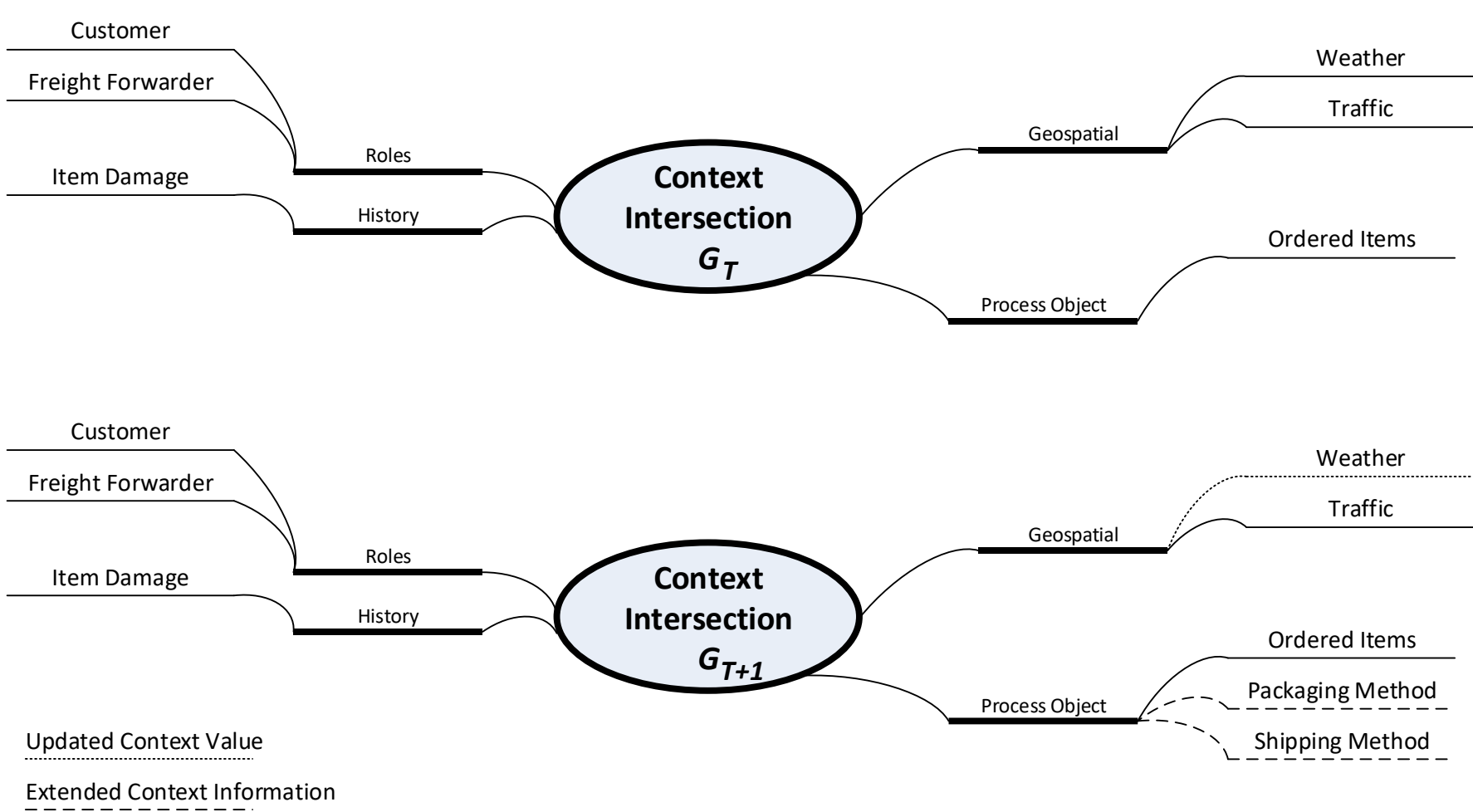

**Fig. 2.** Exemplary Context Intersection at $G_T$ and $G_{T+1}$

This simple context intersection contains elements from three different context categories: geospatial, roles, and process object. The process, freight forwarding, was instantiated with the elements *ordered items, item damage* history*,* and *customer* and *freight forwarder* information as well as the current *traffic* and *weather* information. In the course of the process execution, the *shipping method* and *packaging method* was added. At one point during execution, the weather information was updated. This may have been a routine update or a weather hazard warning. The context engine interprets the current context intersection and propagates the information to the BPM system.

### 3.2 Characteristics

**Hierarchical.** The context intersection provides a context framework and a generic structure. The generic structure enables the context engine to construct the concrete hierarchy for the context categories (cf. Figure 2 for an exemplary hierarchy). We extend the definition of $G$ to represent a hierarchy of context categories.

$$V = V_1 \cup V_2 \cup V_3 \cup \ldots \cup V_k, V_i \cap V_j = \emptyset, i \neq j \ (1)$$

$$e = (v_i, v_j) \in E, i < j \ (2)$$

We further introduce a *predefined context intersection* $C_{pre}$. $C_{pre}$ only includes the set of predefined context categories within the first step of the hierarchy. Therefore, we can conclude:

$$C_{pre} \subseteq V_1 (3)$$

**Extensible.** The structure of a context intersection can change during the execution of an instance $B_P$ of a business process $P$. We suggest that such changes should be limited to adding additional context categories to the current context intersection. This limitation helps to ensure correct execution of the business process as the deletion of context categories within long-running business processes can result in unexpected behavior. It also enables data provenance. For example, packaging information for an ordered item may not be available at the start of the process and will be added to the model later.

We represent changes to a context intersection during the execution of an instance of a business process as a multi-step configuration problem $MP$. The multi-step configuration problem helps to identify valid extensions to the context intersection. $MP_{B_P}$ is defined by a 4-tuple $MP_{B_P} = (MC, K, G_{Start}, G_{End})$, where:

$MC$ is a set of model constraints that must be satisfied in each configuration step.

$K$ is the number of configuration steps during the execution of the instance of the business process model.

$G_{Start}$ is a start configuration of the context intersection at the beginning of the execution of the instance of the business process model. A master context model can serve as a template to create $G_{Start}$. The upper model of Figure 2 can act as a master context model, which is then copied to the context instance model of all new process instances.

$G_{End}$ is the end configuration of the context intersection at the end of the execution of the instance of the business process model.

We define a configuration path from configuration step $T$ up to $N$ configuration steps as a tuple.

$$W_{B_P} = (G_T, G_{T+1}, G_{T+2}, \dots, G_{T+N-2}, G_{T+N-1}, G_{T+N}),\ \ T + N \leq K\ (4)$$

$G_T$ describes the configuration of the context intersection at configuration step $T$. Therefore, $W_{B_P}$ describes changes to the context intersection for $B_P$. The set of constraints $MC$ includes additional constraints besides (1) and (2). We define that only extensions are allowed to a context intersection during run-time, by stating:

$$\forall T \in (0..K-1), V_{i,T} \subseteq V_{i,T+1}, E_T \subseteq E_{T+1} (5)$$

Therefore, the graph $G_T$ is a subgraph of $G_{T+1}$.

**Flexible.** Flexibility refers to how context intersections differ between the executions of different instances of the same business process and the global context (context master model). During the initialization of the execution of a business process instance, a context intersection, which was based on the context master model, can be extended according to the rules presented in this Section. Therefore, the start configuration $G_{Start}$ of a context intersection might differ between two instances of the same business process. For example, not all products are shipped the same way and different context categories may apply for surface or air delivery. Furthermore, the configuration path $B_P$ of the context intersection can differ for two instances of the same business process. Therefore,

$$W_{B_{iP}} \nRightarrow W_{B_{iP}} = W_{B_{jP}}, i \neq j \ (6)$$

holds where $B_{iP}$ and $B_{jP}$ denote the instance $i$ of the business processes $P$ and the instance $j$, respectively.

**Cause-and-effect.** Context categories and their context values are not necessarily independent but often rather inter-dependent. Changes in one context value may affect other context values. Hence, there exist complex interrelationships between context categories, which may be based on aggregation or other (mathematical) functions. They can be relevant to process execution. It may therefore be necessary to specify these relationships in the context models. The resulting changes in context values are similar to derived/ calculated attributes in data models. For example, a delay in the execution of a sub-process such as a partial order delivery may cause an increase in a possible compensation payment.

We refer to a single context value by $v_{id,ts}$ with $id$ denoting a global identifier for the context value and $ts$ denoting a timestamp for the context value, respectively. Over the execution of a business process instance, an existing context value $v_{id,k}$ can be updated with a new context value $v_{id,l}$, $k < l$.

The update of a context value can trigger an automated update of another context value based on a function $f: X \rightarrow Y$. We denote $f$ as cause-and-effect relation, X as the domain of the context value causing the update and Y as the domain of the effected context value. If a context value is effected by multiple cause-and-effect relations, potential update conflicts can arise due to latencies during update propagation. Thus, as a solution, the timestamp of the context value causing the update is propagated.

$$v_{x,k} = f(v_{y,k}), \qquad x \neq y$$

**Time-variant.** Time-variance entails that the usefulness of information relevant to process execution is depending on its and the process's time stamp. Each context information stored in a context category is time-variant. Consequently, there are several conflicts or trade-offs that a context engine needs to process with respect to time. For example, context information for one context category may has different time stamps. New information is preferred over old information as context information depreciates over time.

In addition, information older than a certain time windows and which cannot be refreshed for some reason is associated with a lower level of reliability and importance. It may be of less relevance for extreme actions such as breaking and rolling back of a process.

Furthermore, two context information sources may provide conflicting context information for the same context category (e.g., weather information for a region for the same time window). In this case, additional variables such as reliability (see below) may need to be defined.

**Non-functional attributes.** Lastly, context information may have different non-functional attributes. They can be important factors in the operation of a context-aware BPM system.

For example, reliability or veracity of information may be an issue. It is unlikely that a company will be able to obtain all relevant context value by itself. Hence, trustworthy or certified context providers/ sources with their systems external to the context-aware BPM system are preferred over context information from tertiary or unverified sources. This may result in a tradeoff decision, as context information may come at a charge. Naturally, cheap or free sources are more economical than expensive sources, but may result in delayed or course-granular context information.

# 4 Architecture and Context and Process Adaptation Operations

## 4.1 Architecture

This section presents the software architecture of a system for *context-aware BPM*. Figure 3 shows a BPMN conversation diagram (Object Management Group Inc., 2013), which illustrates the high-level architecture of the system.

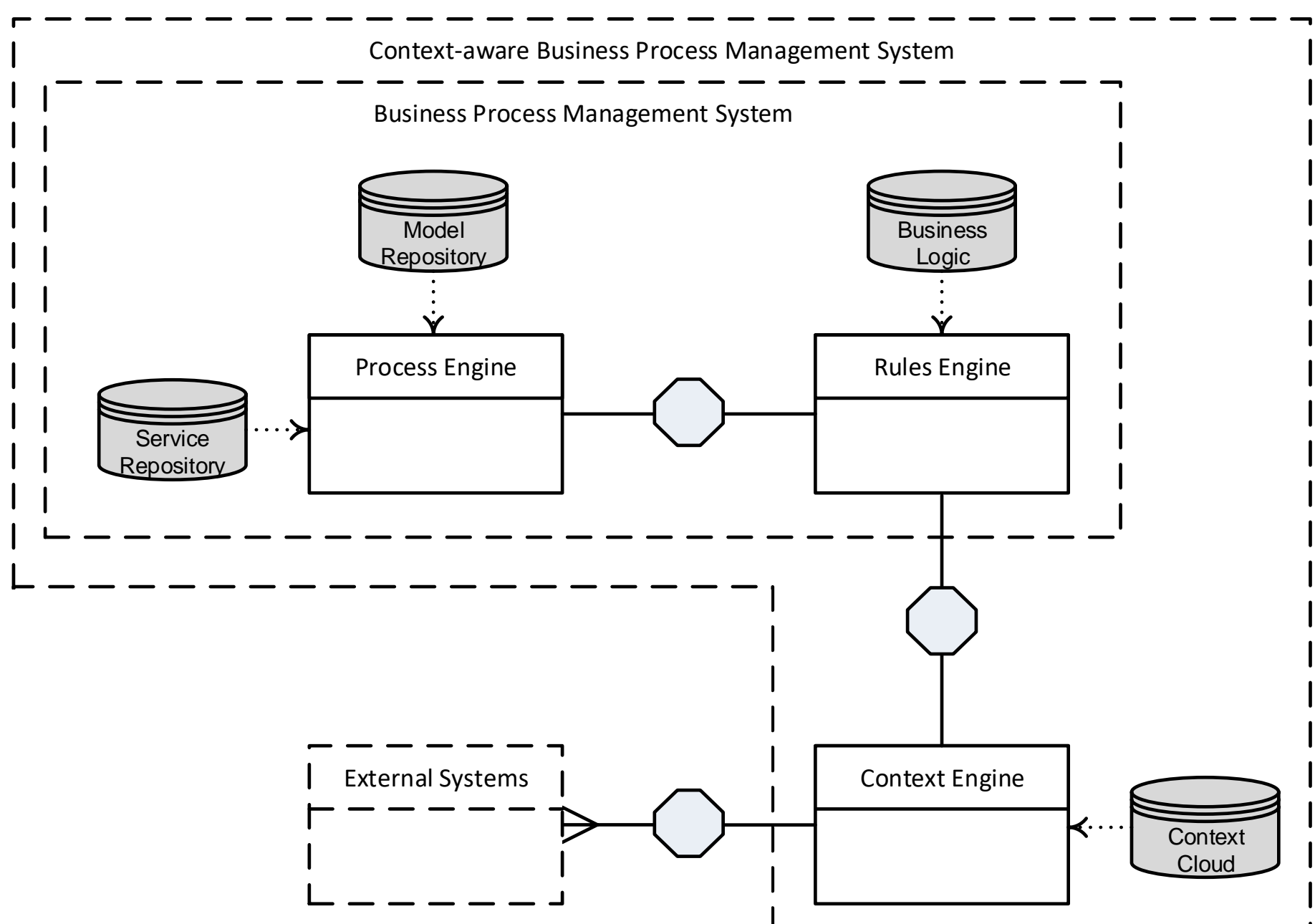

**Fig. 3.** An Architecture for Context-aware BPM

The BPM system comprises a *process engine* and a *rules engine*. The business process engine executes instances of business process models, which are stored within a (*business process) model repository*. Furthermore, the process engine has access to services which it can call, for example via a *service repository*. The process engine can access the rules engine to evaluate rules for decision gates within a business process model. Only the rules engine can interact with a *context engine* to obtain context information. This context information enables an evaluation according to the *business logic* available at decision gates based on relevant, real-time context values. The rules engine can request context values from the context engine. Furthermore, the context engine can push new context information to a registered BPM system. Therefore, the BPM system and context engine communicate in a bidirectional way. The communication process is always started by the BPM system, which notifies the other engines of a new process or process instance.

The context engine uses and administrates the *context cloud* to store context information in context models. The context models and their included context categories determine which context information is directly available to the context engine. If a BPM system requests context information outside the current scope of the context model, the context model can be altered by a context administration agent. The context engine accesses *external systems* to obtain the required context information if it is not available or current in the context cloud. The context administration agent then extends the context model to include the required context categories. External systems can be manifold and depend on the execution context. Apart from private third-party corporate systems such as enterprise systems or supply chain management systems, we consider public information systems for weather, traffic, financial information, but also social media as conceivable external systems.

Furthermore, the context engine polls for updated context information and external systems can push context information to a registered context engine. *External systems* can include systems internal or external to an organization. Examples for internal systems to an organization are customer relationship management systems, enterprise resource planning systems or the BPM system itself. Examples for external systems to an organization are weather information systems or credit reporting systems providing weather information, exchange rate updates, or traffic information.

The context engine could be based on CEP technology as this allows the concurrent observation and processing of a large number of context categories and context update events. Context information can be correlated, aggregated, filtered and, hence, be transformed into higher-level context information which can provide additional benefit to the process execution.

### 4.2 Context Evaluation and Process Adaption

A context aware BPM system has access to several methods for context evaluation and process adaptation. The following selection is non-exhaustive but caters for most use cases.

**Selection of process variants.** Business process models can include alternative process branches for a single business process. Within a business process model, *decision gates* precede process branches. We refer to alternative execution paths represented by processes branches as *process variants*. While executing an instance of a business process model, the BPM system evaluates decision gates to select a process branch. The rules engine evaluates decision gates. However, it is conceivable that decision gates, which represent simple rules (e.g., checking a single value against a threshold), may be evaluated by the process engine itself. Both types of evaluations require context information provided by the context engine. We limit our observations to evaluations by the rules engine only.

**Relevant context.** The evaluation of a decision gate does not necessarily require all context information included in a context model and provided by the context engine for an instance of a business process model. The set of context information that is required to evaluate a decision gate $D$ is called *relevant context* $I_D$. The context engine stores and structures context information within context categories of a context model (cf. previous Section). Therefore, relevant context can be represented by a sub-graph of the context intersection at a configuration step $T$.

$$\mathrm{I_D} \subseteq \mathrm{G_T} \ (7)$$

During the execution of an instance of a business process model, the same decision gate might be evaluated multiple times.

**Threshold-guarded evaluation.** The context engine does not provide means to evaluate the significance of a monitored change in context information. However, the context engine provides notification and updated context information for registered BPM systems. To reduce the number of notifications sent, a BPM system can provide *context notification thresholds* for different types of context information. If the context engine identifies a significant change in relevant context information, the context engine checks specified context notification thresholds. Only if the change is relevant and context information transgresses the specified context notification threshold, a notification is sent to the BPM system.

**(Native) evaluation and re-evaluation.** The evaluation of a decision gate can be triggered in two different ways during the execution of an instance of a business process model. (*Native)* e*valuation* of a decision gate refers to an evaluation according to the process flow. Therefore, multiple evaluations of the same decision gates based on loops within the business process model are considered (native) evaluations. *Re-evaluation* refers to the evaluation of a decision gate exclusively based on change in context information. If a native evaluation of a decision gate depends on context information, the context engine monitors subsequent changes to the used context information. Based on these changes the rules engine can re-evaluate decision gates and trigger appropriate measurements, for example a *break* and/ or *rollback*.

**Break and rollback.** Break and rollback are actions, which can be performed by the process engine. They can only be issued by the rules engine. A break stops the execution of an instance of a business process model. A rollback is the reset of an instance of a business process model to the start or to a previously evaluated decision gate or the start of a process model. Break and rollback can follow an evaluation of a decision gate. In combination with re-evaluations, they allow for the specification of global constraints for a business process. This may include the initialization of compensation processes with or without resorting break and rollback.

# 5 Application and Example

### 5.1 Application

This section describes an archetypical application of the context-aware BPM system from model initialization to process adaptation and completion. Figure 4 shows a BPMN collaboration diagram (Object Management Group Inc., 2013), which illustrates activities and relationships within the architecture presented in Section 4.1. It details Fig.2 and provides a reference process of how the system in the above system architecture interact.

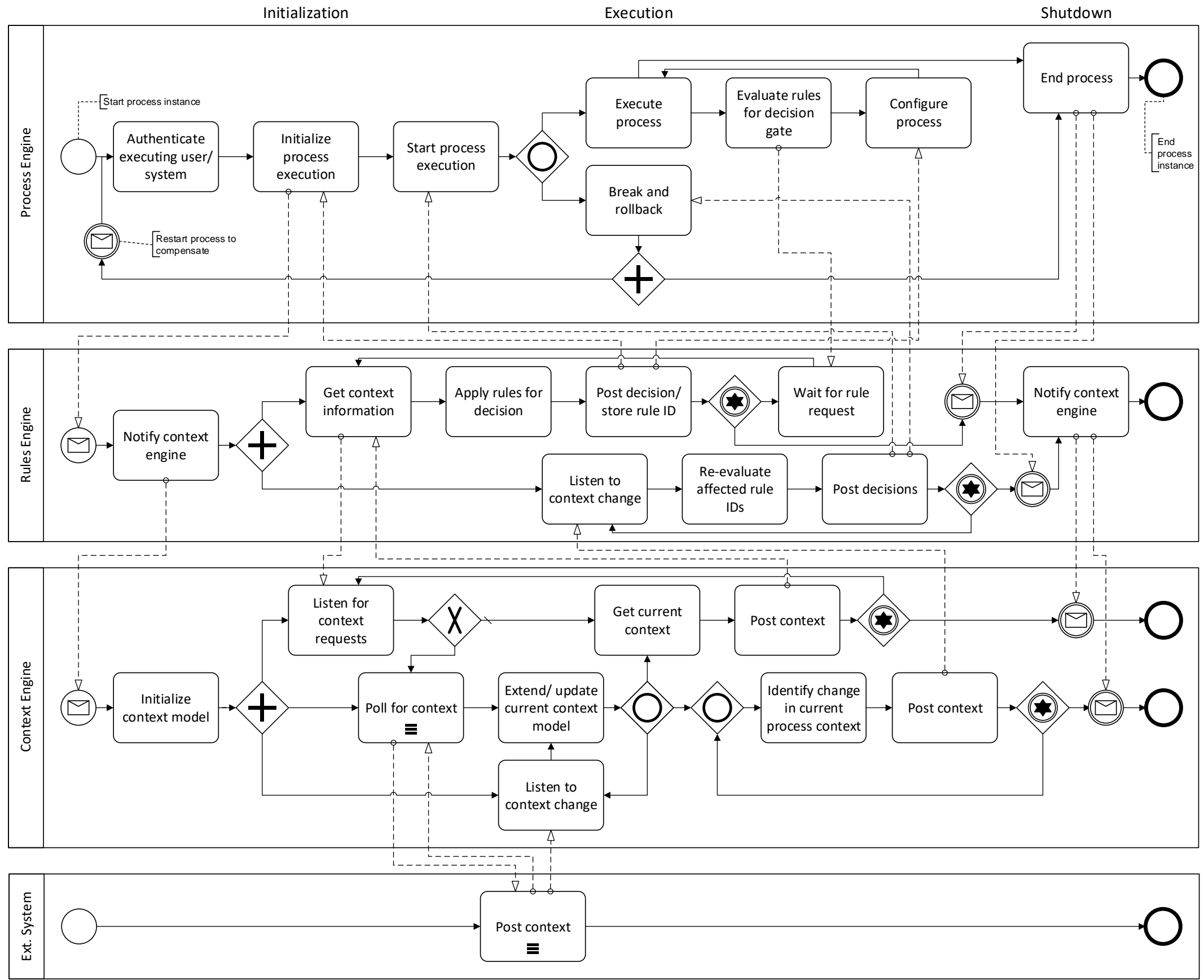

**Fig. 4.** A Reference Process for Context-aware BPM

As introduced in the architecture in the previous Section, the BPM system comprises the process engine and rules engine. However, to better demonstrate messaging interaction on the collaboration level, all engines are modeled in separated pools. Therefore, we use four pools to model process engine, rules engine, context engine, and external systems, which provide context information. Conceptually, the BPM system, which can provide context itself, is regarded as an external system by the context engine in that respect.

The application includes the three phases of *initialization*, *execution*, and *shutdown*. Initialization includes activities required at the start of the execution of a business process instance to configure the build-time model to the current execution context. Execution refers to all activities taking place during the execution of a process instance. Shutdown refers to activities immediately following the completion of a process instance, respectively. It is conceivable to have interactions between the process engine and context engine (cf. also Section 4.2). We do not expand on this for clarity reasons. In the following, we present an archetypical workflow of a context aware process execution in a BPMN collaboration diagram.

**Initialization.** During the initialization phase, the process engine authenticates the executing user or system. In the next step, the process execution is initialized and the current execution context is obtained. The rules engine notifies the context engine to prepare a context model (instance) for the initialized process instance. Furthermore, the rules engine starts to listen to context changes from the context engine relevant to the execution of this process instance. The context engine initializes the context model and begins to listen to context requests from the rules engine. If the current master context model does not include all required information, the context engine polls external systems until all context requirements are satisfied. Furthermore, the context engine starts to listen for context changes based on the context model to keep all values current. If context cannot be published to the subscribed engine, it has to polls for context periodically. Whenever the context engine receives updated or additional context information, the context model is updated or extended. After initialization, the context model is accessed and the context engine posts initial context information to the rules engine. Then it listens for further context value requests. The rules engine evaluates affected rule IDs and configures the initial process variant for execution. Note that at this stage, sub-processes are not initialized but will only be contextualized on a need-to-execute basis. Then the rules engine listens for rule evaluation requests by the process engine. After the process execution is initialized, the process engine starts the execution of the business process instance.

**Execution.** During the execution phase, activities are performed within all four pools. The process engine executes an instance of a business process and listens to break and rollback requests from the rules engine (cf. Section 4.2). If a decision gate is reached during the process execution, the process engine sends a request to the rules engine to start an evaluation and to select the appropriate process variant (cf. Section 4.2). The decision of the rules engine is posted back to the process engine. Based on the decision, the process engine selects the according process variant. Similarly, the initialization of sub-processes can be handled by these tasks.

Furthermore, in extreme cases where context changes make it imperative to cancel the execution of a process as it will not complete or not be economical any more, the process engine can receive a request from the rules engine to break and rollback (cf. Section 4.2).

The rules engine listens for rule requests by the process engine for rule evaluations at decision gates. When a request is received, the rules engine requests the required context values from the context engine and applies the rules connected to the decision gate and posts the decision back to the process engine. It then continues to listen to rule evaluation requests. Rule identifiers associated to the decision are stored to be able to identify relevant decisions taken in case of context change notifications by the context engine. Furthermore, the rules engine listens to changes of relevant context posted by the context engine. If a change is received, a re-evaluation (cf. Section 4.2) of the stored rule IDs is triggered and the decision is forwarded to the process engine. The decision can include the request to break and rollback or to continue process execution as if nothing happened.

The context engine performs three main activities. It listens for context requests from a rules engine, updates and extends the current context model, and identifies changes to the current processes context. If a request from the rules engine is received, the context engine accesses the context model and posts the requested context information back to the rules engine. Accessing the context model might include an extension and/ or update of the current context model. Updates to the context model are received in two ways. First, the context engine can simultaneously poll various external systems for context information. Second, the context engine listens to context changes sent from external systems. If the context engine identifies change in the current context model's values that transgress specified context notification thresholds (cf. Section 4.2), the context engine posts the updated context information to the rules engine, regardless if it was requested or not.

**Shutdown.** If the instance of the business process completes or cancels, the process engine informs the rules engine, which notifies the context engine, respectively. Therefore, the rules engine stops waiting for rule requests from this business process instance from the process engine and does not longer listen to context changes from the context engine. Furthermore, the rules engine notifies the context engine to shut down the of the context model instance. Therefore, the context engine stops listing to context request from the rules engine and stops to extend and update the context model instance. If a context model is being maintained for a process model rather than only for the process instance, this master model may not be affected by the shutdown.

## 5.2 Example

In an example from the logistics sector, a process for the delivery of high-value mining spare parts is being executed by a machine provider. It is crucial for the process that the spare parts reach the destination as a perfect order, which includes (*a*) a correct order entry, (*b*) correct picking, (*c*) delivery in time, (*d*) the item being shipped without damage, and (*e*) invoiced correctly. In this example, we focus on (*c*) and (*d*). Figure 5 depicts the process in multiple simplified BPMN models: a master model, an instantiated instance model, and an instantiated compensation model. The figure does not include the rules engine. An example for a rule statement is included in the description below. Furthermore, the figure does not depict a context model. Figure 2 already supplies all necessary information. It is important to note that though we describe the process for one delivery, it may as well be relevant for all deliveries (i.e., the process model) or a subset of deliveries (e.g., all truck-based deliveries or all deliveries in or to a certain region).

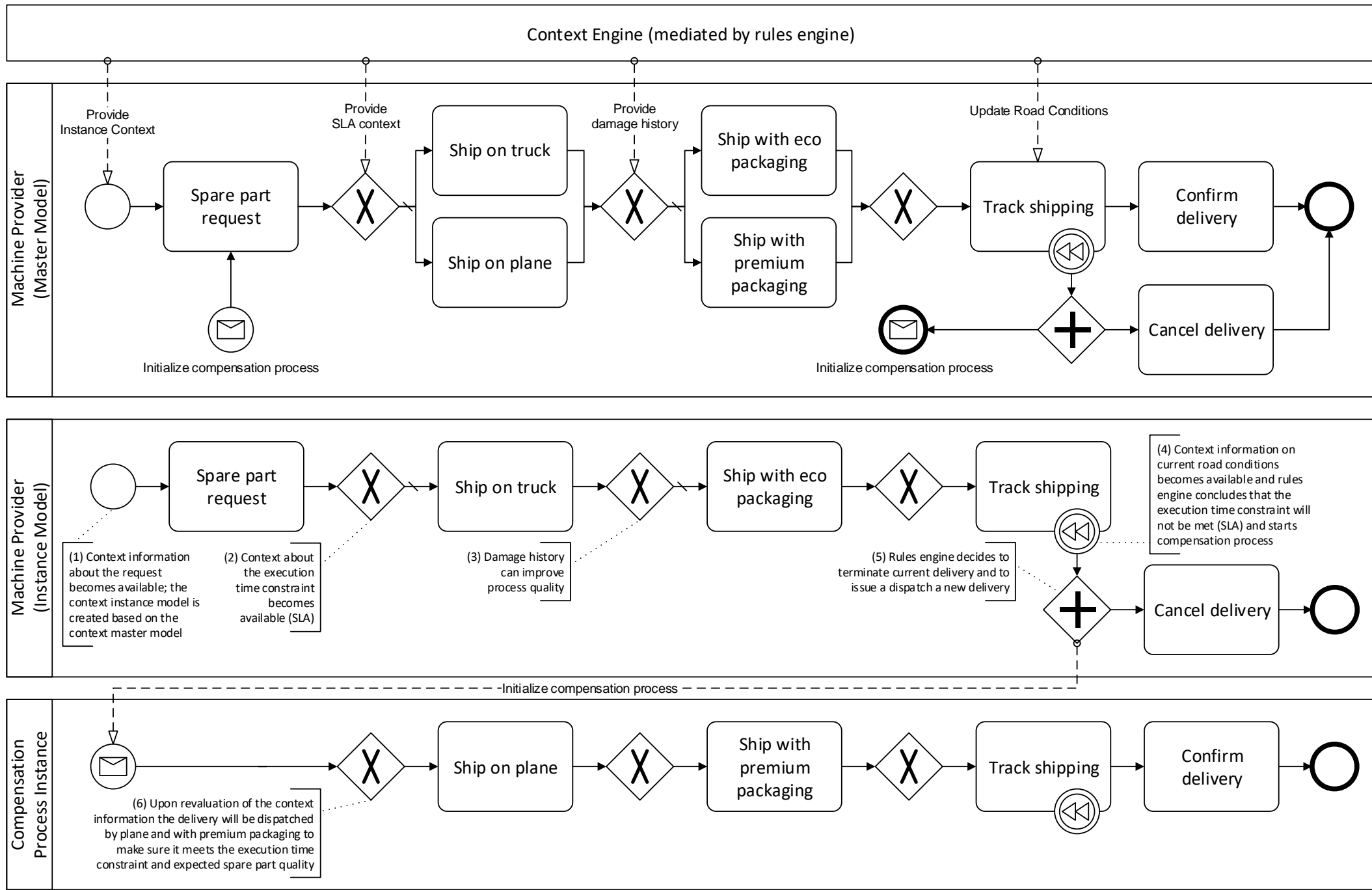

**Fig. 5.** An Exemplary Run-time- and Build-time-Process for Context-aware BPM

The built-time process model (master model) includes variants for different *shipment methods* by truck and by plane. In addition, it includes different *packaging options* (eco and premium) and an *execution time constraint*.

After initialization of run-time process model (instance model) and the context instance model $G_T$ (cf. Figure 2) (1), the process can start. At this stage, the SLA with the customer allows both shipping methods (2). The more economical option of ground delivery is chosen. During process execution, a decision has to be made on the packaging of the spare part. The rules engine queries the context engine for relevant packaging option to the order. Accordingly, the context engine provides the information on the minimal packaging requirements. The rules engine either selects the most economical option or leaves the decision of the relevant alternatives to the user. It may enhance the decision process by providing historical information on customer shipments. If similar goods had been damaged in an economical but theoretically suitable packaging before, it may suggest selecting a more rugged variant. In the case of truck delivery, the economical option is sufficient (3).

After the truck has been loaded and left the warehouse, the context engine notifies the rules engine that the context instance model has updated to $G_{T+1}$ (cf. Figure 2) since heavy thunderstorms have formed en route and have washed away a crucial dirt road making the mining site inaccessible from the truck's direction. The truck would have to be rerouted significantly (4). The rules engine will evaluate this information, calculate a new estimated delivery time, and instruct the BPM system according to rules such as:

```
RULE Delivery SLA
WHEN executionTimeConstraint < estimatedDeliveryTime
     AND maxSLAFineAmount < estimatedSLAFine
THEN start process.compensation.deliveryVariant
END
```

Due to the strict SLA, the rules engine notifies the BPM engine to break and rollback the process and recall the truck as the execution time constraint would be violated (5). At the same time, it initializes a compensation process with the same order data and choses the air delivery variant as SLA can still be met if an identical spare part is dispatched right away. Also, this decision is more economical than paying the fine. Again, packaging options are queried (for air shipment) and the order is dispatched (6). It is to be received as a perfect order from the customer's point of view.

Naturally, as mentioned above this scenario is not limited to adapting the process of a single truck since context instance models can be nested. But it can be extended to all trucks that are impacted by the same context change, for example a massive traffic congestion on a highway since the trucks are all in or moving towards a geospatial coordinate.

This basic scenario already exemplifies some of the core aspects we introduced above. More complex scenarios in different sectors such as banking, urban management, or high value asset management (e.g., oil fields) are conceivable. Not all decisions have to be automated. Naturally, the system can operate at different levels of autonomy (Parasuraman *et al.*, 2000), either taking decisions itself or merely suggesting them.

# 6 Conclusion

### 6.1 Discussion and Limitations

We argue that changes in context such as variation in exchange rates, traffic, or weather can have an impact on the execution of a business process. It can cause slowdowns, even hinder completion, or impact a process's profitability. Most process models are being executed based on static context variables also called workflow relevant data. For long-running processes, it is likely that these static context variables depreciate and are not valid for the duration of the overall process.

Hence, this static assumption is not realistic in a large-scale and inter-organizational process deployment. Therefore, we argue that the execution of a process must be context-aware and, thus, the BPM system should to be accompanied by a context engine, which shadows the process engines process execution and provides updates to the execution context of process instances via a rules engine.

We provide a general system architecture to implement such a system using a master context model and instance context models. This reference architecture assists the full process instance lifecycle: the initialization, execution, and shutdown phase of process execution. Processes can be supplied with an initial context, which is adapted and updated during run-time. Successful or unsuccessful process completions as well as process terminations are handled as well.

The architecture is of generic nature to suit multiple use cases and domains. Therefore, it may lack precision in one domain and it then needs to be contextualized itself. For example, in healthcare there may be other restrictions and requirements than in logistics or life sciences. However, our overall architecture is applicable to all of them and can be adapted to each case. We consciously did not include any domain-specific attributes to keep the architecture applicable to more scenarios.

In literature, there are further context models being discussed. While we propose a context model, models such as those available in formal logic, knowledge representation and reasoning, computational and sociological linguistics, or cognitive psychology should be revisited for concrete implementation. Before use, they need to be reviewed for their dynamic extensibility and the possibility to represent links between master and instance models.

Despite our architecture, the successful application of context-aware BPM is not guaranteed. Context-aware BPM is dependent on current and accurate internal as well as external context information. Thus, companies do not only rely on their own systems for information but also on third-party information providers. These companies' SLA need to be managed carefully as their availability may be an issue and require compensation mechanisms.

Furthermore, context-aware BPM does only provide significant improvements in cost, timeliness, and quality of process execution if decisions due to context can be automated to some respect. Human involvement in decision making is most likely a bottleneck and will slow process adaptation. Context information does not only impact the BPM system but also other associated systems such as the (virtualized) hardware the system is running on. Here, the knowledge of context change can be used to adapt the run-time environment as well, for example to provide more resources to the BPM system to accelerate task execution and, thus, overall process performance. This entails that context communication channels need to be open and flexible (yet secure) enough to allow for the registration of further system at the rule or context engine.

Finally, context-aware BPM might not be the right solution for all types of processes as running and maintaining a context engine in conjunction with a BPM system will require adequate skills and cause additional cost. Ultimately, the value of information timeliness plays a similar role as in the decision for or against real-time systems (Olsson and Janiesch, 2015).

### 6.2 Summary and Outlook

In this paper, we presented an architecture for context-aware BPM consisting of a BPM system, comprising a process engine, a rules engine, and context engine, which collects and analyzes context information from external systems. We also describe the workflow of a typical context-aware process execution and give an example from the logistics sector. The example is supposed to illustrate the capability with a comprehensible example. Naturally, high volume processes with automatable lower value decisions are a primary candidate for this system architecture proposal.

Future work will focus on the sophistication of the context model as well as the concretization of the architecture components. Furthermore, there is a need to define integrated modeling languages for analytic and procedural tasks (cf. also blinded) as well as interaction patterns for context engines und BPM systems to understand better the impact that context change can have on process instances. Furthermore, we plan to integrate context engines with novel tracing systems to populate context models with short latencies (blinded). In the domain of smart grids, we currently evaluate an integration of context engines with different context systems, for example weather forecast and smart meter gateways.

# References

Adams, M., ter Hofstede, A., Russell, N., and van der Aalst, W. (2010), "Dynamic and Context-Aware Process Adaptation", in: Wang, M. and Sun, Z. (Eds.), *Handbook of Research on Complex Dynamic Process Management: Techniques for Adaptability in Turbulent Environments*, IGI Global, Hershey, PA, pp. 104-136.

Adams, M., ter Hofstede, A.H.M., van der Aalst, W.M.P., and Edmond, D. (2007), "Dynamic, Extensible and Context-Aware Exception Handling for Workflows", in *OTM Confederated International Conferences On the Move to Meaningful Internet Systems (OTM). Lecture Notes in Computer Science Vol. 4803, Vilamoura*, Springer, pp. 95-112.

Anastassiu, M., Santoro, F.M., Recker, J., and Rosemann, M. (2016), "The Quest for Organizational Flexibility: Driving Changes in Business Processes Through the Identification of Relevant Context", *Business Process Management Journal*, Vol. 22, No. 4, pp. 763-790.

Arló-Costa, H.L. (1999), "Epistemic Context, Defeasible Inference and Conversational Implicature", in Bouquet, P., Serafini, L., Brézillon, P., Benerecetti, M., and Castellani, F. (Eds.), *Second International and Interdisciplinary Conference on Modeling and Using Context (CONTEXT). Lecture Notes in Computer Science Vol. 1688, Trento*, Springer, pp. 15-27.

Becker, J., Kugeler, M., and Rosemann, M. (2011), *Process Management: A Guide for the Design of Business Processes*, 2nd Edn., Springer, Berlin.

Bettini, C., Brdiczka, O., Henricksen, K., Indulska, J., Nicklas, D., Ranganathanf, A., and Riboni, D. (2010), "A Survey of Context Modelling and Reasoning Techniques", *Pervasive and Mobile Computing*, Vol. 6, No. 2, pp. 161-180.

Brézillon, P., Pomerol, J.C., and Saker, I. (1998), "Contextual and Contextualized Knowledge: An Application in Subway Control", *International Journal of Human-Computer Studies*, Vol. 48, No. 3, pp. 357-373.

Chen, G. and Kotz, D. (2000), "A Survey of Context-aware Mobile Computing Research", Dartmouth Computer Science Technical Report No. TR2000-381, Hanover, NH.

Cipriani, N., Wieland, M., Groímann, M., and Nicklas, D. (2011), "Tool Support for the Design and Management of Context Models", *Information Systems*, Vol. 36, No. 1, pp. 99-114.

Clark, H. and Carlson, T. (1981), "Context for Comprehension", in: Long, J. and Baddeley, A. (Eds.), *Attention and Performance*, Vol. IX, Lawrence Erlbaum Associates, Hillsdale, NJ, pp. 313-330.

Crozier, J. (2006), *Collins Essential English Dictionary*, 2nd Edn., Collins, Glasgow.

Delfmann, P., Janiesch, C., Knackstedt, R., Rieke, T., and Seidel, S. (2006), "Towards Tool Support for Configurative Reference Modeling: Experiences from a Meta Modeling Teaching Case", in Brockmanns, S., Jung, J., and Sure, Y. (Eds.), *2nd Workshop on Meta-Modelling and Ontologies (WoMM). Lecture Notes in Informatics Vol. 96, Karlsruhe*, pp. 61-83.

Dey, A.K. (2001), "Understanding and Using Context", *Personal and Ubiquitous Computing*, Vol. 5, No. 1, pp. 4-7.

Dey, A.K. and Abowd, G.D. (2000), "Towards a Better Understanding of Context and Context-Awareness", in *CHI Workshop on The What, Who, Where, When, and How of Context-Awareness, The Hague*, pp. 1-12.

Etzion, O. and Niblett, P. (2011), *Event Processing in Action*, Manning Publications, Cincinnati, OH.

Gu, T., Pung, H., and Zhang, D. (2005), "A Service-oriented Middleware for Building Context-Aware Services", *Journal of Network and Computer Applications*, Vol. 1, No. 28, pp. 1-18.

Hallerbach, A., Bauer, T., and Reichert, M. (2008), "Context-based Configuration of Process Variants", in *3rd International Workshop on Technologies for Context-aware Business Process Management (TCoB), Barcelona*, pp. 31-40.

Halliday, M.A.K. (1978), *Language as Social Semiotic: The Social Interpretation of Language and Meaning*, Edward Arnold, London.

Herzberg, N., Meyer, A., and Weske, M. (2015), "Improving Business Process Intelligence by Observing Object State Transitions", *Data & Knowledge Engineering*, Vol. 98, pp. 144-164.

Janiesch, C., Matzner, M., and Müller, O. (2011), "A Blueprint for Event-driven Business Activity Management", in Rinderle, S., Toumani, F., and Wolf, K. (Eds.), *9th International Conference on Buisness Process Management (BPM). Lecture Notes in Computer Science Vol. 6896, Clermont-Ferrand*, Springer, pp. 17-28.

Janiesch, C., Matzner, M., and Müller, O. (2012), "Beyond Process Monitoring: A Proof-of-Concept of Event-driven Business Activity Management", *Business Process Management Journal*, Vol. 18, No. 4, pp. 625-643.

Kokinov, B. (1999), "Dynamics and Automaticity of Context: A Cognitive Modeling Approach", in Bouquet, P., Serafini, L., Brézillon, P., Benerecetti, M., and Castellani, F. (Eds.), *Second International and Interdisciplinary Conference on Modeling and Using Context (CONTEXT). Lecture Notes in Computer Science Vol. 1688, Trento*, Springer, pp. 200-213.

Krumeich, J., Weis, B., Werth, D., and Loos, P. (2014), "Event-Driven Business Process Management: Where are we now? A Comprehensive Synthesis and Analysis of Literature", *Business Process Management Journal*, Vol. 20, No. 4, pp. 615-633.

Kunz, S., Fickinger, T., Prescher, J., and Spengler, K. (2010), "Managing Complex Event Processes with Business Process Modeling Notation", in Mendling, J., Weidlich, M., and Weske, M. (Eds.), *2nd International Workshop on Buisness Process Modeling Notation (BPMN). Lecture Notes in Buisness Information Processing Vol. 67, Potsdam*, Springer, pp. 78-90.

Luckham, D. (2002), *The Power of Events: An Introduction to Complex Event Processing in Distributed Enterprise Systems*, Addison-Wesley Professional, Boston, MA.

Michelberger, B., Mutschler, B., and Reichert, M. (2012), "A Context Framework for Process-oriented Information Logistics", in *15th International Conference on Business*

*Information Systems (BIS). Lecture Notes in Business Information Processing Vol. 115, Vilnius*, pp. 260-271.

Möhring, M., Schmidt, R., Härting, R.-C., Bär, F., and Zimmermann, A. (2014), "Classification Framework for Context Data from Business Processes", in F., F. and Mendling, J. (Eds.), *10th BPM Workshop on Social and Human Aspects of Business Process Management (BPMS2). Lecture Notes in Business Information Processing Vol. 202, Eindhoven*, Springer, pp. 440-445.

Mundbrod, N., Grambow, G., Kolb, J., and Reichert, M. (2015), "Context-Aware Process Injection: Enhancing Process Flexibility by Late Extension of Process Instances", in *OTM Confederated International Conferences "On the Move to Meaningful Internet Systems" (OTM). Lecture Notes in Computer Science Vol. 9415, Rhodes*, Springer, pp. 127-145.

Nunes, V.T., Santoro, F.M., Werner, C.M.L., and Ralha, C.G. (2016), "Context and Planning for Dynamic Adaptation in PAIS", in Reichert, M. and Reijers, H.A. (Eds.), *4th International BPM Workshop on Decision Mining & Modeling for Business Processes (DeMiMoP). Lecture Notes in Business Information Processing Vol. 256, Innsbruck*, Springer, pp. 471-483.

Object Management Group Inc. (2013): "Business Process Model and Notation (BPMN) Version 2.0.2", available at: http://www.omg.org/spec/BPMN/2.0.2/PDF (accessed 2017-09-06).

Olsson, L. and Janiesch, C. (2015), "Real-time Business Intelligence und Action Distance: Ein konzeptionelles Framework zur Auswahl von BI-Software", in *12. Internationalen Tagung Wirtschaftsinformatik (WI), Osnabrück*, pp. 691-705.

Parasuraman, R., Sheridan, T., and Wickens, C. (2000), "A Model for Types and Levels of Human Interaction with Automation", *IEEE Transactions on Systems, Man and Cybernetics - Part A: Systems and Humans*, Vol. 30, No. 3, pp. 286-297.

Ploesser, K. (2013), "A Design Theory for Context-Aware Information Systems", Queensland University of Technology, Brisbane.

Ploesser, K., Recker, J., and Rosemann, M. (2010), "Building a Methodology for Context-aware Business Processes: Insights from an Exploratory Case Study", in Johnson, R. and de Villiers, C. (Eds.), *18th European Conference on Information Systems (ECIS), Pretoria*, pp. 1-12.

Rosemann, M., Recker, J., Flender, C., and Ansell, P. (2006), "Understanding Context-Awareness in Business Process Design", in Spencer, S. and Jenkins, A. (Eds.), *17th Australasian Conference on Information Systems (ACIS), Adelaide*, pp. 1-10.

Rosemann, M. and van der Aalst, W.M.P. (2007), "A Configurable Reference Modelling Language", *Information Systems*, Vol. 32, No. 1, pp. 1-23.

Russell, N., van der Aalst, W., and ter Hofstede, A. (2006) "Workflow Exception Patterns", 18th International Conference Advanced Information Systems Engineering (CAiSE). Lecture Notes in Computer Science Luxembourg, Springer, pp. 288-302.

Saidani, O. and Nurcan, S. (2009), "Context-Awareness for Adequate Business Process Modelling", in Flory, A. and Collard, M. (Eds.), *3rd IEEE International Conference on Research Challenges in Information Science (RCIS), Fez*, pp. 177-186.

Serral, E., de Smedt, J., and Vanthienen, J. (2014), "Extending CPN Tools with Ontologies to Support the Management of Context-Adaptive Business Processes", in F., F. and Mendling, J. (Eds.), *3rd BPM Workshop on Data- & Artifact-centric BPM (DAB). Lecture Notes in Business Information Processing Vol. 202, Eindhoven*, Springer, pp. 198-209.

Sowa, J.F. (1999), *Knowledge Representation: Logical, Philosophical, and Computational Foundations*, Brooks Cole Publishing, Pacific Grove, CA.

Stettner, K. and Janiesch, C. (2009), "Key Requirements for a Context-aware Service Marketplace: An Expert's Perspective", in Scheepers, H. and Davern, M. (Eds.), *20th Australasian Conference on Information Systems (ACIS), Melbourne*, pp. 164-173.

UN/CEFACT (2009): "UN/CEFACT Core Components Technical Specification. Version 3.0", available at: https://www.unece.org/fileadmin/DAM/cefact/codesfortrade/CCTS/CCTS-Version3.pdf (accessed 2017-09-06).

von Ammon, R., Ertlmaier, T., Etzion, O., Kofman, A., and Paulus, T. (2009), "Integrating Complex Events for Collaborating and Dynamically Changing Business Processes", in *2nd Workshop on Monitoring, Adaptation and Beyond (MONA+), Stockholm*, pp. 1-16.

von Halle, B. (2001), *Business Rules Applied: Business Better Systems Using the Business Rules Approach*, John Wiley & Sons, New York, NY.

Wieland, M., Kopp, O., Nicklas, D., and Leymann, F. (2007), "Towards Context-aware Workflows", in: Pernici, B. and Gulla, J.A. (Eds.), *CAiSE'07 19th International*

*Conference on Advanced Information Systems Engineering: Trondheim, 11-15 June, 2007: proceedings of the workshops and doctoral consortium* Vol. 2, Tapir Acasemic Press, Trondheim, pp. 577-591.

Wieland, M., Nicklas, D., and Leymann, F. (2011), "Benefits of Business Process Context for Human Task Management", *International Journal of Trade, Economics and Finance*, Vol. 2, No. 4, pp. 304-311.

Workflow Management Coalition (1999): "Terminology and Glossary", available at: http://www.wfmc.org/standards/docs/TC-1011_term_glossary_v3.pdf (accessed 2017-09-06).

World Wide Web Consortium (2006): "Delivery Context Overview for Device Independence. W3C Working Group Note 20 March 2006", available at: http://www.w3.org/TR/di-dco/ (accessed 2017-09-06).

zur Muehlen, M. and Indulska, M. (2010), "Modeling Languages for Business Processes and Business Rules: A Representational Analysis", *Information Systems*, Vol. 35, No. 4, pp. 379-390.